\documentclass[aps,pre,twocolumn,pacs]{revtex4}
\usepackage{amsmath,bm,epsfig}
\usepackage{lscape}

\newcommand{\C}[1]{{\mathcal{#1}}}

\usepackage[latin1]{inputenc}

\begin{document}
\title{Fronts Propagation at the Onset of Plastic Yielding}
\author{Eran Bouchbinder$^{1,2}$ and Ting-Shek Lo$^{1,3}$}
\affiliation{$^1$Dept. of Chemical Physics, Weizmann Institute of Science, Rehovot 76100, Israel,\\
$^2$Racah Institute of Physics, Hebrew University of Jerusalem, Jerusalem 91904, Israel,\\
$^3$Dept. of Physics, The Chinese University of Hong Kong, Shatin, Hong Kong}
\begin{abstract}
The existence of a finite threshold, the yield stress, for the
onset of plastic yielding is a universal feature of plasticity. This jamming-unjamming transition is naturally
accounted for by the dynamics of a bistable internal state field.
We show, within the athermal Shear Transformation Zones (STZ)
theory of amorphous plasticity, that the transition is accompanied by the propagation of
plastic fronts. We further
show that the mean-field theory cannot select the velocity of these
fronts and go beyond the mean-field description to include fluctuations and correlations effects, resulting in new nonlocal
terms in the equations. Finally, we demonstrate that the new terms, with an associated intrinsic lengthscale, provide a velocity selection mechanism for the plastic fronts.
\end{abstract}
\maketitle

\section{Introduction}

A complete understanding of the dynamics of plastic deformation in
low temperature, or athermal, amorphous systems remains a major
theoretical challenge in statistical and condensed matter physics.
The response of these systems to the application of external driving
forces exhibits some universal features like a transition from
jamming to flow at the yield stress, orientational memory
effects of recent deformation and strain localization \cite{06MR}.
In this work we focus on the spatiotemporal patterns associated
with the transition from jamming to flow at the onset of plastic
yielding, in the framework of the recently developed athermal Shear
Transformation Zones (STZ) theory of amorphous systems \cite{98FL, 07BLPa}. This transition is controlled by the applied stress. For stresses below a material dependent threshold, the yield stress, these systems exhibit elastic deformation as well as transient plastic flow that vanishes at a finite time. The resulting state is ``jammed'' in the sense that it carries a finite stress without flowing.
On the other hand, for stresses that exceed the yield stress, these systems exhibit persistent plastic flow, implying some remarkable ``unjamming'' (or yielding) transition at the yield stress.

The main difficulty in developing a theory of amorphous plasticity, that among other things should predict plastic yielding at a finite yield stress, is that the identity and
characteristics of the microstructural objects that ``carry''
plastic deformation in amorphous (disordered) systems remain quite
elusive. However, a growing body of experimental and simulational
evidence seems to point towards a unifying picture in which low
temperature amorphous plasticity involves stress driven
configurational rearrangements of localized regions composed of a
small collection of the relevant elementary entities (e.g.
particles, molecules, grains, colloids, bubbles) \cite{STZ}. The
existence of such a unifying picture, which is independent of the
elementary entities and their microscopic interactions, is fully
consistent with the existence of universal features mentioned above.

In line with this accumulating evidence, the athermal Shear-Transformation-Zones (STZ) theory of amorphous plasticity
views STZ as localized groups of particles that are more
susceptible to shearing transformation under stress than their
surroundings. Upon surpassing a local threshold an STZ can undergo
a finite irreversible shear in a given direction. Once
transformed, due to a local redistribution of stresses, an STZ
resists further deformation in that direction, but is particulary
sensitive to reverse shear deformation if the local applied stress
changes sign. Therefore, an STZ is conceived as a two-state
system that can transform between its internal states depending on
the magnitude and direction of the local stress. In addition, the
stress redistribution that accompanies an STZ transition
can induce the creation and annihilation of other STZ at a rate
proportional to the local energy dissipation (recall
that thermal fluctuations are assumed to be absent or negligible). The minimal
two-state assumption, along with the deformation driven creation
and annihilation of STZ, provides a mechanism for retaining and losing orientational memory of
previous deformation and is sufficient to capture the transition from
jamming to flow at a finite yield stress.

We choose to study the spatiotemporal characteristics of the unjamming transition using STZ theory for two major reasons. First, the existence of the transition is described in this theory, as will be explained in detail below, by an exchange of dynamic stability in the equations of motion for an internal state field. We find such a description superior to theoretical approaches that simply postulate the existence of a threshold stress above which plastic flow persists \cite{Lubliner}. More generally, we find an approach that aims at identifying proper order parameters, or internal state fields, and writing equations of motion for them based on microscopic insights as well as symmetries and conservation laws very appealing \cite{98FL, 07BLPa}.
Second, recent work has shown that the STZ equations capture a variety of phenomena observed in computer simulations and laboratory experiments \cite{03ELP,07SKLF,07BLPb,07MLC,07LM,08Bouch,08BLP,07LKXOD,08FTL}. Most relevant for our purposes here are the results reported in \cite{07LKXOD,08FTL}. In \cite{07LKXOD}, the two-state nature of STZ has been observed directly in bubble-raft experiments
and two-dimensional foam simulations. It was found that groups of bubbles can make transitions, the so-called T1 events that are realizations of STZ transitions in foam, between two states if the direction of the applied force is reversed shortly after an event occurs.
However, if the force is not reversed until after
other events have occurred nearby, then the orientational memory is
lost, implying that an STZ is annihilated. In \cite{08FTL} the STZ equations were shown to agree quantitatively with transient shear reversal behavior of a granular flow in a Taylor-Couette cell. This result provides the first direct support to an equation that will be central in the analysis to follow.

We proceed by introducing the relevant mean-field STZ equations in Sect. \ref{STZ}, where we also discuss the existence of the transition from jamming to flow at a finite yield stress and the physics behind it. In Sect. \ref{fronts} we focus on the spatiotemporal characteristics of the unjamming transition. We show that there exist propagating front solutions if the equations are modified to include effects beyond mean-field as well as an intrinsic lengthscale. Section \ref{sum} offers some concluding remarks.

\section{Elements of the Athermal Shear
Transformation Zones (STZ) Theory}
\label{STZ}

Consider a two-dimensional athermal
amorphous system under the application of a pure shear stress $s$. The two states of an
STZ are oriented along the principal axes x (the direction of the
force) and y (the perpendicular direction), and denoted by
``$\pm$''. The number densities of STZ in the ``$\pm$'' states are
denoted by $n_\pm$. The configurational disorder of the system is
characterized by an effective disorder temperature $\chi$ that is
assumed to govern the total density of STZ, with $n_\infty
e^{-1/\chi}$ being the steady state density of STZ under
persistent deformation. $\chi$ was shown very recently to play an
important role in the deformation dynamics of amorphous solids
\cite{07SKLF,07BLPb,07MLC,07LM,08Bouch}; however, its dynamics in the present context are not
of prime interest and will not be discussed. The internal state
fields $n_\pm$ and $\chi$, in addition to the applied stress $s$,
determine the macroscopic plastic rate of deformation $D^{pl}$
according to \cite{07BLPa}
\begin{eqnarray}
\label{Dplav}
 \tau_0 D^{pl}\!\!\!&=&\!\!\!\lambda \Bigl(\!R(s)n_-\!-\!R(-s)n_+\!\Bigr)\ , \\
 \label{ndot2} \tau_0\,\dot n_{\pm}\!\!\!&=&\!\!\!R(\pm s)n_{\mp}\!-\!R(\mp
s)\,n_{\pm}\!+\! \Gamma\!\left(\!\frac{n_{\infty}e^{-1/\chi}}{
2}\!-\!n_{\pm}\!\right)\!.
\end{eqnarray}
Here $\lambda$ is a material-specific parameter with the
dimensions of area. $\tau_0$ is the basic time scale and $R(\pm
s)/\tau_0$ are the rates for forward and backward transitions
respectively. The athermal condition implies that $R(s)$ vanishes
for $s\!<\!0$, i.e. there are no transitions in the direction
opposite to the direction of the applied force \cite{07BLPa}.

Eq. (\ref{Dplav}) simply states that the macroscopic plastic rate of deformation $D^{pl}$ results from STZ transitions between their two internal states, depending on the driving force through $R(s)$ and on the population of STZ in each of the two states.
Eqs. (\ref{ndot2}) are Master equations for the population densities $n_\pm$ themselves. The first two terms on the right hand side describes the number conserving process in which STZ change their state from ``+'' to ``-'' or vice versa. The third term on the right hand side is a coupling term that accounts for interactions between STZ. As mention above, the
stress redistribution that accompanies an STZ transition
can induce the creation and annihilation of other STZ . This is a crucial aspect of the theory.
The rate of STZ creation $\Gamma$, which is a positive definite scalar, is assumed to be proportional to the rate of plastic dissipation $2sD^{pl}$ \cite{98FL, 07BLPa}. For low temperature we obtain \cite{98FL, 07BLPa}
\begin{equation}
\label{Gamma}
\Gamma=\frac{2\tau_0 s D^{pl} }{s_y \epsilon_0 e^{-1/\chi}} \ ,
\end{equation}
where $s_y$ is a material-specific parameter of stress dimension, to be shown below to be equal the yield stress.
$s_y$ has an interesting physical meaning: it quantifies how much of the dissipated energy is invested in creating
new STZ. When it is large, only a small amount of the dissipated energy is invested in creating new STZ (available for irreversible transitions) and vice versa. This interpretation immediately suggests that $s_y$ is related to the yield stress; see \cite{08Lan} for a discussion of this issue from an entropic point of view.

We proceed by setting the total STZ density to its steady state value,
$n_+\!+\!n_-\!=\!n_\infty e^{-1/\chi}$ and by defining an orientational order parameter $m$ (which is generally a tensor, see \cite{07MLC}) as
\begin{equation}
m\equiv \frac{n_+-n_-}{n_+ +n_-} \ .
\end{equation}
This internal state field represents the bias of the STZ populations between the $\pm$ states.
With these choices we can rewrite Eqs. (\ref{Dplav})-(\ref{ndot2}) as
\begin{eqnarray}
\label{eq:Dpl0}
\tau_0 {D^{pl}} &=& \epsilon_0 e^{-1/\chi} {\cal C}(s)\left[{\rm sgn}\left(\frac{s}{s_y}\right)-m\right] \ , \\
\label{eq:m0}  \tau_0 \dot m &=& 2{\cal C}(s)\left[{\rm sgn}\left(\frac{s}{s_y}\right)-m\right]\left(1-\frac{s~m}{s_y}\right) \ .
\end{eqnarray}
Here $\epsilon_0\!\equiv\!\lambda n_\infty$ and
\begin{equation}
\C C(s)\equiv\frac{R(s)+R(-s)}{2} \ .
\end{equation}
The theoretical framework in which the amorphous system is characterized by
coarse-grained internal state fields, in addition to stress and
strain, is a major point of divergence compared to other approaches \cite{98FL}
and plays a crucial role in the discussion below, where the field
$m$ is the central object.

In \cite{08FTL}, Eqs. (\ref{eq:Dpl0})-(\ref{eq:m0}) were directly applied to predict transient granular flow in experiments of shear reversal in a Taylor-Couette cell. The theory was shown to agree well with the experimental data, encouraging us to further study it in the present context.
The first issue to be discussed is the stable steady state
solutions of Eq. (\ref{eq:m0}). It has two stationary solutions:
\begin{enumerate}
\item A jammed state with $m =\pm 1$ (depending on the sign of
$s$), where all the STZ are either in the ``+'' or the ``-'' state
respectively, and $D^{pl}= 0$ in Eq. (\ref{eq:Dpl0}),
\item A flowing state with $m=s_y/s$ and $D^{pl}\! \ne\! 0$ in Eq. (\ref{eq:Dpl0}).
\end{enumerate}
 These two solutions coincide
at $s/s_y\! =\! \pm 1$. It is straightforward to show that the
jammed state is dynamically stable for $|s|\! \le \!s_y$, while the
flowing state is dynamically stable for $|s|\! >\! s_y$. An exchange
of stability occurs at $s\! =\! s_y$, which indeed shows explicitly that $s_y$ is the yield stress.
The fact that $s_y$ was introduced in Eq. (\ref{Gamma}) as a proportionality constant between the rate of STZ creation $\Gamma$ and the rate of energy dissipation $2sD^{pl}$ is yet another point of divergence with respect to standard approaches. In STZ theory the yield stress $s_y$, as was explained before, is intimately related to the ability of the material to create new STZ as a result of plastic dissipation, i.e. STZ transitions in nearby locations \cite{08Lan}. In this interpretation, a material can yield, i.e. support steady plastic flow, if enough STZ per unit time are created due to other STZ transitions. If this rate is not sufficiently large, then existing STZ are exhausted in transitions at a given direction and not enough new STZ are available for additional transitions that are needed for attaining a steady state. For a given applied stress, whether or not $\Gamma$ is large enough, is determined by $s_y$. This is the essence of yielding in STZ theory.

As a result, the onset of
plastic yielding at a finite threshold is a {\em dynamic}
phenomenon manifested as an exchange of stability of the bistable
field $m$ \cite{98FL}. These results have similar implications for
strain-rate $\dot\gamma$ controlled experiments (complementary to
the stress-controlled experiments discussed up to now) where the
existence of a finite yield stress means that a steady state with
a finite stress is obtained in the limit $\dot\gamma\to 0$. The
existence of a dynamic threshold is a signature of an underlying
bistability (or multistability), which is a natural way to get a
finite dissipation at a vanishing strain-rate. Note
that all these conclusions are entirely independent of the
material function ${\cal C}(s)$. We thus propose that the
bistability of the orientational order parameter $m$ is a crucial
feature of the theory of amorphous plasticity.

\section{The transition from Jamming to Flow}
\label{fronts}

We now proceed to analyze the spatiotemporal characteristics of this
jamming-unjamming transition. We substitute Eq. (\ref{eq:Dpl0}) into
Eq. (\ref{eq:m0}) to obtain
\begin{equation}
\label{mEq} \tau_0 \dot m = 2~{\cal C}( s)(1- m)\left(1-
\frac{\displaystyle m s}{\displaystyle s_y}\right)\ ,
\end{equation}
where we have assumed $s\!>\!0$. Consider then an experiment in
which the stress is ramped to a constant value smaller than the
yield stress (i.e. $s\!<\!s_y$). The system is assumed to be initially
undeformed and isotropic such that the orientational order parameter
satisfies $m(t\!=\!0)\!=\!0$. Eqs. (\ref{eq:Dpl0}) and (\ref{mEq})
then predict that sub-yield plastic deformation occurs as $m$
relaxes to the jamming fixed-point $m\!=\!1$ on a typical time scale
$\tau_{jam}(s)$
\begin{equation}
\label{relax} \tau_{jam}(s) \simeq \frac{\tau_0}{2~{\cal C}(s)\left(1-
\frac{\displaystyle s}{\displaystyle s_y}\right)}\ ,
\end{equation}
during which $D^{pl}\!\to\! 0$. The jamming time scale
$\tau_{jam}(s)$ diverges as $s\!\to\!s_y^-$, i.e. approaching the
yield stress from below, and as $s\!\to\!0^+$. The latter
divergence is a result of the athermal condition, leading to
${\cal C}(s\!\to\!0)\!\to\!0$ \cite{07BLPa}. Therefore, Eq.
(\ref{relax}) predicts that jamming is obtained by progressively
slower creep-like sub-yield deformation as $s$ approaches either 0
or 1. Moreover, it provides a way to measure the phenomenological
function ${\cal C}(s)$ in the range where the jamming time is
experimentally (or simulationally) accessible.

Suppose now that $s/s_y$ is ${\cal O}(1)$, but not very close to unity
such that the system becomes jammed on a realistic time scale;
then suppose that after jamming the applied stress is ramped again
to a value above the yield stress, i.e. $s\!>\!s_y$. The system is
now in the jammed state, $m\!=\!1$, that is unstable for
$s\!>\!s_y$. In these situations we generally expect fluctuations to
propagate as fronts, converting an unstable state into a stable
one. Therefore, we are looking for translational invariant
solutions of the form $m(x,t)\!=\!m(x-ut)$; here we assume that
the x-dimension is much larger than the y-dimension and that the
front propagates from left to right. Our task now is to solve Eq.
(\ref{mEq}) for this ansatz, with the boundary conditions
\begin{equation}
\label{BC}
 m\left(x\!\to\! -\infty\right) \!\to\! \frac{s_y}{s},\quad\quad
 m(x\!\to\! +\infty) \!\to\!1 \ .
\end{equation}
The first boundary condition corresponds to the flowing state left
behind the front, while the second one corresponds to the jammed
state ahead of it. A solution can be readily obtained, yielding
\begin{widetext}
\begin{equation}
\label{m_front} m(x,t) = \frac{\displaystyle 1-\left(\frac{s_y^2}{
s^2}-\frac{s_y}{s} \right) \exp\left[{\frac{-2 \C
C(s)(s/s_y-1)(x-ut)}{\tau_0 u}}\right]}{\displaystyle 1-\left(\frac{s_y}{
s}-1 \right)\exp\left[{\frac{-2 \C C(s)(s/s_y-1)(x-ut)}{\tau_0
u}}\right]}\ .
\end{equation}
\end{widetext}
This seems to be a propagating front solution in which a plastically
deforming region, $D^{pl}\!\ne\!0$, invades a jammed region, $D^{pl}\!=\!0$.

However, Eq. (\ref{m_front}) is not a unique solution of the problem
since it is valid for {\em any} velocity $u$. More importantly, a
velocity $u$ cannot be selected by Eq. (\ref{mEq}) {\em in
principle} since in the absence of an intrinsic lengthscale a
velocity cannot be {\em dimensionally} constructed. Therefore, if
we aim at describing the spatiotemporal patterns that accompany
the onset of plastic flow, we cannot avoid addressing the
fundamental problem of missing a lengthscale in our theory. In
fact, two lengthscales are already implied within our theoretical
framework; a finite STZ density implies a typical distance between
STZ, which can be thought of as a microstructural correlation
length. This lengthscale might be relevant as different STZ can
interact via the stress and displacement fluctuations generated
when an STZ transforms between its internal states. An evidence for the existence of such a lengthscale was given, for example, in \cite{05SF,Glass}; it can be identified with the typical size of areas of quasi-crystal-like short-range order of \cite{05SF} or with the typical size of regions free of liquid-like defects of \cite{Glass}.

The range in which these interactions can affect STZ transitions also implies a
lengthscale, possibly larger than the previous one, that is
determined by a combined effect of the magnitude of the stress
fluctuation, its range and the distribution of STZ transition
thresholds. This lengthscale characterizes the scale in which the
probability of finding a sufficiently large stress perturbation
(that can overcome the local transition threshold) is not
exponentially small. Therefore, we should consider {\em nonlocal} stress fluctuations that can induce STZ transitions at different locations where STZ already exist and are close to their transition threshold. This effect was stressed by several authors \cite{94BA, 02BVR, 07LC} and was assumed to be the origin of the cascades of rearrangements observed in athermal quasi-static simulations \cite{04ML}. With this physical intuition in mind, we
should extend the homogeneous and local mean-field theory to
include some inhomogeneous and nonlocal effects.

In summary, our theory should incorporate nonlocal terms that
account for the following physical effects: (i) The stress and
displacement fluctuations that are generated by STZ transitions,
(ii) The joint spatial probability distribution function of STZ
that includes a correlation length, (iii) The STZ transition
thresholds distribution. Therefore, we add to the left hand side
of Eqs. (\ref{Dplav})-(\ref{ndot2}) terms that are weighted
integrals over existing terms that include the transition rates
$R(\pm s)$; these new terms represent that idea that STZ
transitions from ``+'' states to ``-'' states (or vice versa) at a
given location result in a properly weighted change in the rate of
transitions at different locations. Rearranging the modified
equations we arrive at the following equation for $m(x,t)$
\begin{widetext}
\begin{equation}
\label{m_modified} \tau_0 \dot m(x,t) =2~{\cal C}(s)\left[1- m(x,t)\right]\left[1- \frac{s~m(x,t)}{s_y} \right]+
2~\C C(s) \left[1-\frac{s~m(x,t)}{s_y} \right]\!\! \int_{-\infty}^{\infty}\!\!\!\!K(x-x')\left[1-m(x',t)\right]dx'. 
\end{equation}
\end{widetext}
The phenomenological kernel $K(x)$ in Eq. (\ref{m_modified}),
whose dimension is $(length)^{-1}$, represents the physics of STZ
correlations and interactions discussed above. The fact that it is
a one-dimensional function of x alone already incorporates several
approximations. First, there is some evidence that STZ transitions generate quadrupolar elastic fields \cite{94BA}. Here this anisotropic
structure is neglected due to the assumption that the y-dimension
is much smaller than the x-dimension. Second, the kernel $K$ might
include some time dependence that accounts, for example, for the
wave nature of the stress and displacement fluctuations. We assume
that this time-dependence can be integrated out without affecting
the basic physics we are trying to describe. In choosing a specific functional form for the kernel $K(x)$ we do not want to make
any definite claim as to whether the incorporated lengthscale is a short-distance correlation length or a longer-distance cutoff scale discussed above, but to stress that our theoretical considerations {\em entails the existence of a lengthscale}.
Both physical possibilities share the feature that the immobile STZ interact via nonlocal stress fluctuations which can be schematically described by
\begin{equation}
\label{kernel} K(x) =
\frac{\alpha}{\ell}\exp\left(-\frac{x^2}{2\ell^2}\right) \ ,
\end{equation}
where $\ell$ is the characteristic lengthscale and $\alpha$ is
the amplitude. Note that we do not consider here very
long-range kernels, i.e. with power-law tails, as we believe that the STZ transition thresholds distribution limits the range in which stress fluctuations can induce STZ transitions, as explained above.

Eq. (\ref{m_modified}), with the kernel of Eq. (\ref{kernel}), poses a non-trivial front selection problem. As in many other front selection problems \cite{PhysRep}, there might exists a family of propagating front solutions from which a unique solution is selected dynamically. A complete analysis of Eq. (\ref{m_modified}) is well beyond the scope of the present work. However, we aim here at demonstrating that the proposed new terms indeed lead to front selection. To that aim, we first consider the possibility that the nonlocal contribution is a small correction to the mean-field equation and treat Eq. (\ref{m_modified}) perturbatively with $\alpha$ being the small parameter in the expansion. We thus write
$m=m^{(0)}+\alpha m^{(1)}+{\cal O}(\alpha^2)$, where $m^{(0)}$ is given in Eq. (\ref{m_front}).
The first order equation in $\alpha$ selects the velocity, from which we estimate
\begin{equation}
\label{velocity}
u \simeq \frac{2~\ell~ \C C(s)}{\displaystyle \tau_0} \left(\displaystyle \frac{s}{s_y}-1\right)\ .
\end{equation}
It is important to note that the nonlocal term is a singular perturbation in the sense that there is no solution
at all in the limit $\alpha\!\to\!0$.

\begin{figure}
\centering
\epsfig{width=.5\textwidth,file=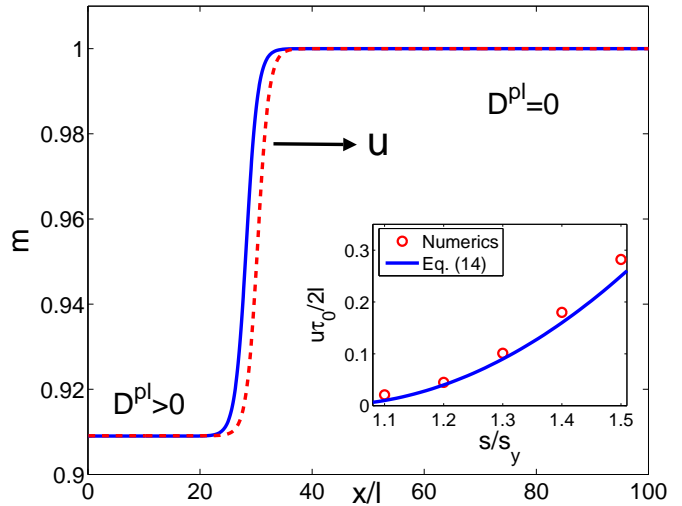}
\caption{(Color online) The propagating $m$ front solution of Eq. (\ref{m_modified}) with $\alpha=0.01$ and $s=1.1$ (solid line) compared to the analytic prediction given by Eq. (\ref{m_front}) (dashed line), shifted by a few time units for clarity. (Inset) The velocity $u$ (in units of $2\ell/\tau_0$) as a function of $s$, both for the numerics (open circles) and the prediction of Eq. (\ref{velocity}) (solid line).}\label{front}
\end{figure}

As we cannot offer a similar analysis for the general case where $\alpha$ is not necessarily small, we studied Eq. (\ref{m_modified}) numerically by
choosing $s\!>\!s_y$ and introducing the homogeneous jammed state,
$m\!=\!1$, with various localized perturbations.
In the numerical calculations we choose, for simplicity, $\C C(s)\!=\!H(s-s_y)(s-s_y)$ for $s\!>\!0$. Here $H(\cdot)$ is the Heaviside unit step function.
We have found that all perturbations converge, for sufficiently large times, to
a {\em unique} front profile with a {\em unique} velocity $u$.
This result shows that indeed the new nonlocal term in Eq.
(\ref{m_modified}) provides a front velocity and profile selection
mechanism for the problem at hand. In Fig. \ref{front} we show the
numerical front profile for a small amplitude $\alpha$ and the
analytic prediction of Eq. (\ref{m_front}) (shifted by a few time
units for clarity). In the inset we
compare the numerical results for the velocity $u$ as a function
of $s$ to the prediction of Eq. (\ref{velocity}). We observe that in this regime, i.e. for small
$\alpha$, the profile is, to a very good approximation, given by
Eq. (\ref{m_front}) with the selected velocity satisfying Eq.
(\ref{velocity}). Upon increasing the amplitude $\alpha$, the
nonlocal term becomes more dominant and the solution of Eq.
(\ref{m_front}) is no longer accurate.
%%%%%%% FIGURE 1 %%%%%%%%%%%%%%%%%%
%%%%%%%%%%%%%%%%%%%%%%%%%%%%%%%%%%%
We thus conclude that in the presence of the nonlocal interaction term, with the associated lengthscale $\ell$, the proposed theory predicts the existence of propagating front solutions that remain to be observed in computer simulations and experiments.

\section{Concluding Remarks}
\label{sum}

In this paper we asked what are the spatiotemporal characteristics of the transition between a jammed state found at applied stresses below the yield stress $s_y$ and a homogeneously plastic flow state found at applied stresses above the yield stress $s_y$. In our opinion, this is a fundamental question in the field of plasticity of amorphous systems. We addressed this problem in the framework of the recently developed athermal Shear-Transformation-Zones (STZ) theory, which in our opinion offers a promising route for developing a predictive theory of amorphous plasticity.
This theory describes the phenomenon of plastic yielding at a finite threshold in terms of an exchange of dynamic stability in the equation of motion for an orientational order parameter $m$. Our
main result is that this theory predicts the existence of
propagating front solutions that accompany the yielding transition provided that nonlocal effects are introduced.
The new nonlocal terms, that account for nonlocal STZ interactions, incorporate a lengthscale that is missing in the original theory.
Our predictions for the existence of plastic fronts at the yielding transition should be tested in computer simulations and experiments.
This may provide further support for the analytic structure of the developing STZ theory,
substantiate the existence of the bistable orientational order parameter $m$ and shed some light on the missing intrinsic lengthscale.
As the existence of a yield stress is a fundamental property of materials, elucidating the spatiotemporal nature of the yielding transition is of major importance.

{\bf Acknowledgements} E. B. acknowledges support from the Horowitz Center for Complexity Science and the Lady Davis Trust.

\end{document}